# Generating Ten BCI Commands Using Four Simple Motor Imageries

Nuri Korhan, Tamer Ölmez and Zümray Dokur

Istanbul Technical University
Department of Electronics and Communication Engineering, Istanbul, Turkey
corresponding author: korhan@itu.edu.tr

**ABSTRACT —** The brain computer interface (BCI) systems are utilized for transferring information among humans and computers by analyzing electroencephalogram (EEG) recordings. The process of mentally previewing a motor movement without generating the corporal output can be described as motor imagery (MI). In this emerging research field, the number of commands is also limited in relation to the number of MI tasks; in the current literature, mostly two or four commands (classes) are studied. As a solution to this problem, it is recommended to use mental tasks as well as MI tasks. Unfortunately, the use of this approach reduces the classification performance of MI EEG signals. The fMRI analyses show that the resources in the brain associated with the motor imagery can be activated independently. It is assumed that the brain activity induced by the MI of the combination of body parts corresponds to the superposition of the activities generated during each body part's simple MI. In this study, in order to create more than four BCI commands, we suggest to generate combined MI EEG signals artificially by using left hand, right hand, tongue, and feet motor imageries in pairs. A maximum of ten different BCI commands can be generated by using four motor imageries in pairs. We observe in the literature that the classification performances are adversely affected as the number of classes is increased, and the success rates for the MI EEG signals with more than four classes are poor. This study aims to achieve high classification performances for BCI commands produced from four motor imageries by implementing a small-sized deep neural network (DNN). The presented method is evaluated on the four-class datasets of BCI Competitions III and IV, and an average classification performance of 81.8% is achieved for ten classes. The above assumption is also validated on a different dataset which consists of simple and combined MI EEG signals acquired in real-time. Trained with the artificially generated combined MI EEG signals, DivFE resulted in an average of 76.5% success rate for the combined MI EEG signals acquired in real-time.

**Keywords**

EEG
Combined motor imagery
Multi-class EEG classification
Brain-computer interface
Convolutional neural network

## 1. Introduction

In order to establish an artificial intelligence based means for human-computer interaction, the EEG signals acquired by a BCI system can be translated into commands that perform a specific desired action. There is a growing interest on the researches carried out in the field of BCI applications in recent years. Especially for the classification of MI EEG signals many new methods have been proposed. The process of mentally previewing a movement without any physical output can be described as motor imagery. The activation of the specific areas on the human brain is similar when he/she imagines it or performs the action in real world.

The EEG signals mostly resemble random noise. Unlike the electrocardiogram, they do not have any specific wave pattern. When working with MI EEG signals, one should know that (*i*) the EEG signals are of low spatial resolution, (*ii*) the EEGs show variations between sessions and subjects, (*iii*) classification of many different classes with high success rates it is not so easy, (*iv*) data collection can sometimes be difficult if the subject needs a pre-training/preparation to be trained accordingly, and (*v*) in order to compensate all those issues a complex algorithm might be needed to be developed in order to increase the classification accuracy, which, in turn, makes the system very hard to implement in real life. The number of electrodes placed on the subject's head determines the spatial resolution. The resolution can be improved by increasing the number of electrodes keeping in mind the fact that it will require more computational power for the BCI system. Besides, using many electrodes causes the BCI experiments to be impractical for the researchers and uncomfortable for the subjects. One of the greatest challenges of BCI is that the manifestation of the same imagined movement in the EEG recordings varies between the subjects, even between the two different sessions for the same subject. Another disadvantage of working with EEG recordings is that, the number of BCI commands is limited with the number of MI tasks that can be discriminated from each other; in the current literature, mostly two or four commands (classes) are studied. One classifier which shows good performance for one subject might not be able to show the similar level of performance for another one, which is also another cause of problem. Therefore, the training of the subjects before the final data collection task becomes very critical. To make the final system efficient, not only the correct classification but also the speed is also an important factor for the classification of MI EEG signals. Since the researchers are mostly focused on the classification accuracy using the benchmark datasets [1, 2] for the comparisons with the literature, more complicated classifier structures have been developed which require higher computation power and time [3–32]. Such complicated and power consuming algorithms make it impossible to integrate them into portable embedded systems. Moreover, in order to determine the optimum features specific for a subject, a relatively large amount of data is required, but the datasets used in the literature are of small sizes [1, 2].

In a previous study, researchers suggested the use of a novel data augmentation method on the small-sized datasets [33], and also proposed the use of a novel convolutional neural network (CNN) structure, called as DivFE [33, 34], in order to improve the classification performance of MI EEG signals. In this study, while dealing with the problems stated in (*iii–v*), we will be more concerned with increasing the number of BCI commands.

Another problem is to increase the number of BCI commands which is limited with the number of MI tasks that can be recognized individually. In the current literature, four commands (tongue, feet, left and right hands) are studied at most. In Fig. 1, motor imageries for the left and right feet are somewhere inside the red ellipses. MI EEG signals for the left and right feet were not even used in all BCI Competition datasets. The reason for not using left foot and right foot as different classes is that it is almost impossible to discriminate MI EEGs of left and right feet from each other. For this reason, both feet are used as a single class in these benchmark datasets [1, 2]. MI centers for left and right feet are inside the brain and are very close to each other, as seen in the Fig. 1. For this reason, in four-class problems, classes were generally chosen as left hand, right hand, tongue and feet. The MI locations of these four classes have been chosen particularly far from each other. In fact, it is necessary to use more electrodes on the scalp in order to distinguish close centers in the brain. In the literature, there is a study that succeeds in distinguishing MIs related to the left foot and right foot by increasing the number of electrodes on the scalp [35]. However, due to the increased complexity and computational burden, this approach makes the system less efficient.

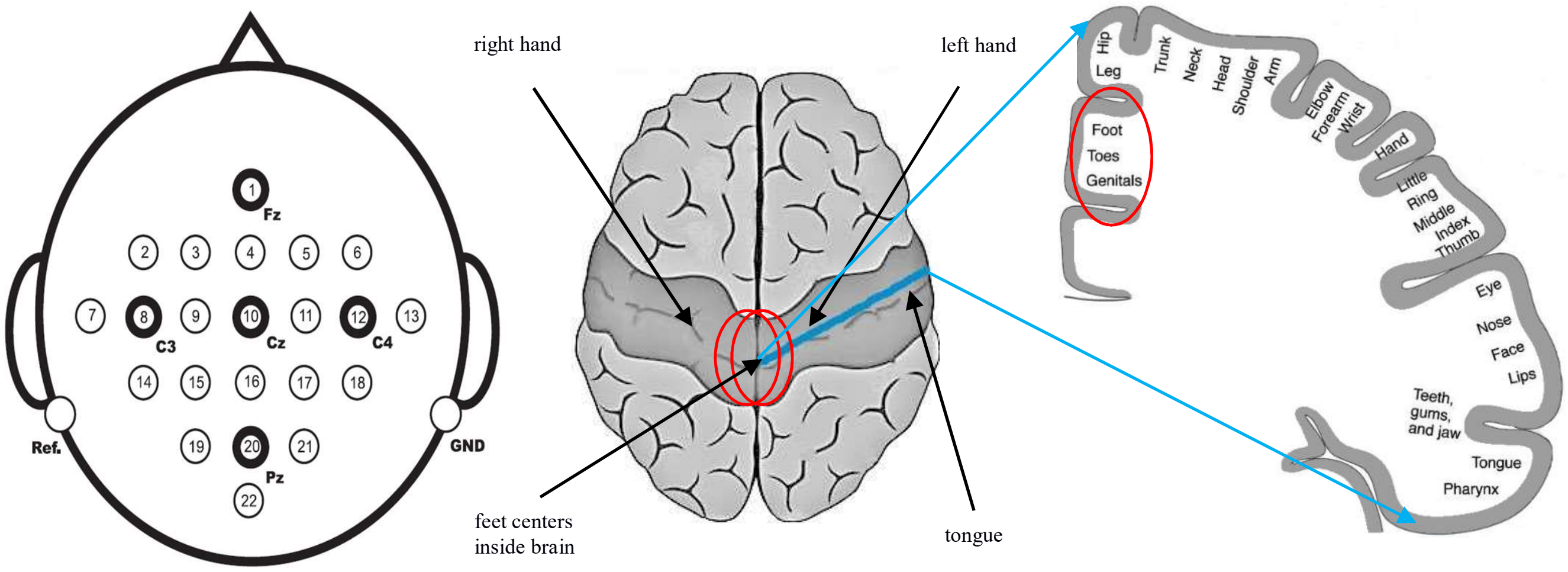

**Fig. 1** Locations of the left hand, right hand, feet, and tongue centers in the human brain.

Another way to increase the number of BCI commands is to use the MIs in pairs [36–41]. In [36], EEG signals from six right-handed healthy subjects were recorded by using the OpenViBE platform at a sampling frequency of 256 Hz. The EEG cap that incorporates 26 electrodes was placed according to the international 10-20 system. The EEG recordings corresponding to the left hand, right hand, both hands, and a ‘rest’ task on which the subjects must not think about any motor movement, were classified with an average success rate of 51.6%. In the study [38], eight right-handed healthy subjects (six females and two males, 23–25 years old) participated. For the EEG based MI-BCI control, 20 electrodes (labeling F3, FZ, F4, FC3, FCZ, FC4, C5, C3, C1, CZ, C2, C4, C6, CP3, CPZ, CP4, P5, P1, P2, and P6) were mounted in both horizontal and vertical directions over the motor cortex. The MI EEG signals corresponding to the left hand, right hand, both hands and feet were classified with an average success rate of 54.2% by using the common spatial patterns (CSP) method. In the study [40], the researchers had made experiments with seven subjects that were given eight different MI tasks (rest, left hand, right hand, feet, left hand & feet, right hand & feet, both hands & feet, and both hands) which they had to perform in a series of trials. The EEG cap was comprised of 26 electrodes, namely, Fp1, Fpz, Fp2, Fz, FC5, FC3, FC1, FCz, FC2, FC4, FC6, C5, C3, C1, Cz, C2, C4, C6, CP5, CP3, CP1, CPz, CP2, CP4, CP6, and Pz, and the electrodes were located over the extended international 10-20 system positions to cover the cerebral cortex. Eight different MI EEG signals were classified with an average success rate of 55% by using the CSP. In [41], seven female and three male subjects (23–25 years old, all right-handed) were given three simple MI tasks, three combined MI tasks, and a ‘rest’ task. In the experiments, the EEG recordings were collected via 64 Ag/AgCl electrodes placed in compliance with the international 10-20 system. The MI EEG signals corresponding to the left hand, right hand, feet, both hands, left hand combined with right foot, right hand combined with left foot, and rest were classified with an average success rate of 70% by using the CSP.

In the studies of [36–41], it is observed that CSP gives low classification performance. One of the reasons for the low performance is that the CSP method is used in the combined MIs. The CSP is a transformation that reveals the distinction between the samples of the different classes. In simple MIs, event-related desynchronization and synchronization (ERD/ERS) signals are generated at different locations in the brain. However, in combined MIs, such as the left hand versus both hands, the ERD/ERS signals will be generated at the locations that are close to each other in the brain. Therefore, the use of the CSP may not be convenient in distinguishing the combined MIs.

In another study [36], the subjects were given the tasks of performing four different MIs: left hand, right hand, both hands, and rest. In the study, Figs. 2 and 3 show ERD/ERS ratios of MIs corresponding to the left hand, right hand, both hands, and rest for C3 and C4 electrodes, respectively. As a result, the ERD/ERS ratios of combined MI signals (both hands) appear as a superposition of simple MI signals (left hand or right hand). Fig. 4 in [36] further shows that this assumption is correct. It is a result of the subjects' being able to make several motor movements at the same time.

In this study, we suggest the use of tongue, feet, left hand, and right hand MIs in pairs to create more than four BCI commands by using the four-class datasets of the BCI Competition III and IV. Maximum of ten (4 simple MI + C(4,2) combined MI) different BCI commands are generated artificially by using four simple MIs in pairs. In the literature, it is observed that the classification performance generally decreases as the number of classes increases, and success rates for the MI EEG signals with more than four classes are poor. The aim of this study is to achieve high classification performance for ten BCI commands that are artificially produced by combining four simple MIs. In this context, ERD/ERS ratios for simple and combined MI signals are analyzed, and it is investigated how to generate combined MI signals artificially using the superposition of simple MI signals. In the Computer Simulations section, the assumption developed in this study will be validated on a different dataset which consists of simple and combined MI EEG signals.

## 2. METHODOLOGY

### 2.1 Structure of the proposed convolutional neural networks

Most of the DNN structures are composed of convolutional layers which extract features from the samples, and fully connected layers which pave the path to the classification. In addition to these layers a DNN structure contains pooling and batch normalization layers to increase the performance of the network. Another indispensable component of a DNN is the layers of activation units that determine the behavior of the neurons and this layer is usually selected as rectified linear units (ReLU).

Having a number of learning filters with small receptive fields that are convolved with each channel of the EEG recordings in order to create one-dimensional activation maps, the critical part of the CNN is the convolutional layers. The main purpose of these layers is to reveal features that will be used to create other features that are of greater importance in representing the data. When the size of the representation needs to be decreased, either the strides of the filters are increased or the pooling layers are inserted between the convolutional layers in order to carry out the down-sampling operation while also contributing to the translation invariance (reducing the variance of the data). In comparison to the other functions such as average-pooling and min-pooling, max-pooling is the most favored nonlinear function to implement pooling operation in research and applications. Utilization of pooling layers in the CNNs in literature is wide despite the fact that the need is problem dependent. Likewise, whether a pooling layer should be placed after a convolutional layer is mostly determined by trial and error method. Pooling layers are deployed between all convolutional layers of the model. In this model, the activation function is selected as ReLU, $f(x)$=max(0, $x$), which performs the nonlinear operation of neutralizing all the negative values of the activation maps. Additionally, two of the most common regularization methods are used to obtain a more robust model. Having all the components of both the feature extractor (CNN) and the classifier (fully connected neural network-FCNN) at hand, we face the challenging task of tuning all hyper-parameters such as the number of hidden layers in the network, the number of neurons in the hidden layers, the number of convolutional layers, and the size and number of filters in the convolutional layers, simultaneously. This tuning process may take weeks in the studies that contain a relatively large dataset. If the DNN is constructed by using only CNN part of the aforementioned structure, the number of hyper-parameters that needs to be tuned decreases dramatically. In [33, 34], researchers put forward a DNN structure that contains the CNN, and replaced the FCNN with a minimum distance network (MDN). This structure is called DivFE (Divergence-based Feature Extractor) and shown in Fig. 2 in detail.

When the FCNN is used in a network that aims to solve a machine learning problem it brings a number of issues that are not easy to tackle. Some of these issues are high memory requirement, high computational cost, difficulty of convergence, and additional hyper-parameters. An MDN, on the other hand, deals with extracted features in a much easier manner. When the MDN is used as a classifier, the class labels of the MDN are set to the Walsh vectors (columns or rows of the Walsh matrix). The label of the data is determined by the distance between the feature extractor's (FE) output and the given previously determined Walsh vector. The MDN node that is closest to the output of the FE sets the class label for that data point. Given that, $M$ is the dimension of the Walsh matrix (which is equal to 16 in this study), $O_j$ is the $j$th output of the flatten layer (see Fig. 2), and $H_{k,j}$ is the $j$th element of the Walsh vector belonging to the $k$th class, the class label is determined according to the following equation:

$$\boldsymbol{D}_k = \sum_{j=1}^{M} (O_j - H_{k,j})^2 \qquad \boldsymbol{D}_i = \min_k(\boldsymbol{D}_k) \tag{1}$$

where index $i$ is the class label of the data point that is determined by the MDN.

### 2.2 Training process of the convolutional layers

In deep learning applications, as the number of parameters to be trained increases, training the network becomes more and more difficult. The reason for this phenomenon is that as the number of trainable parameters increases, the number of local minima also increases and the algorithm becomes more likely to stuck in some local minima. It has been shown in previous studies [33, 34] that feature extractor and classifier can be trained separately, one after another, as long as the feature extractor is trained before the classifier. Once the features are proven to be extracted successfully, the fully connected layers can be replaced with a minimum distance classifier as it is done in this study.

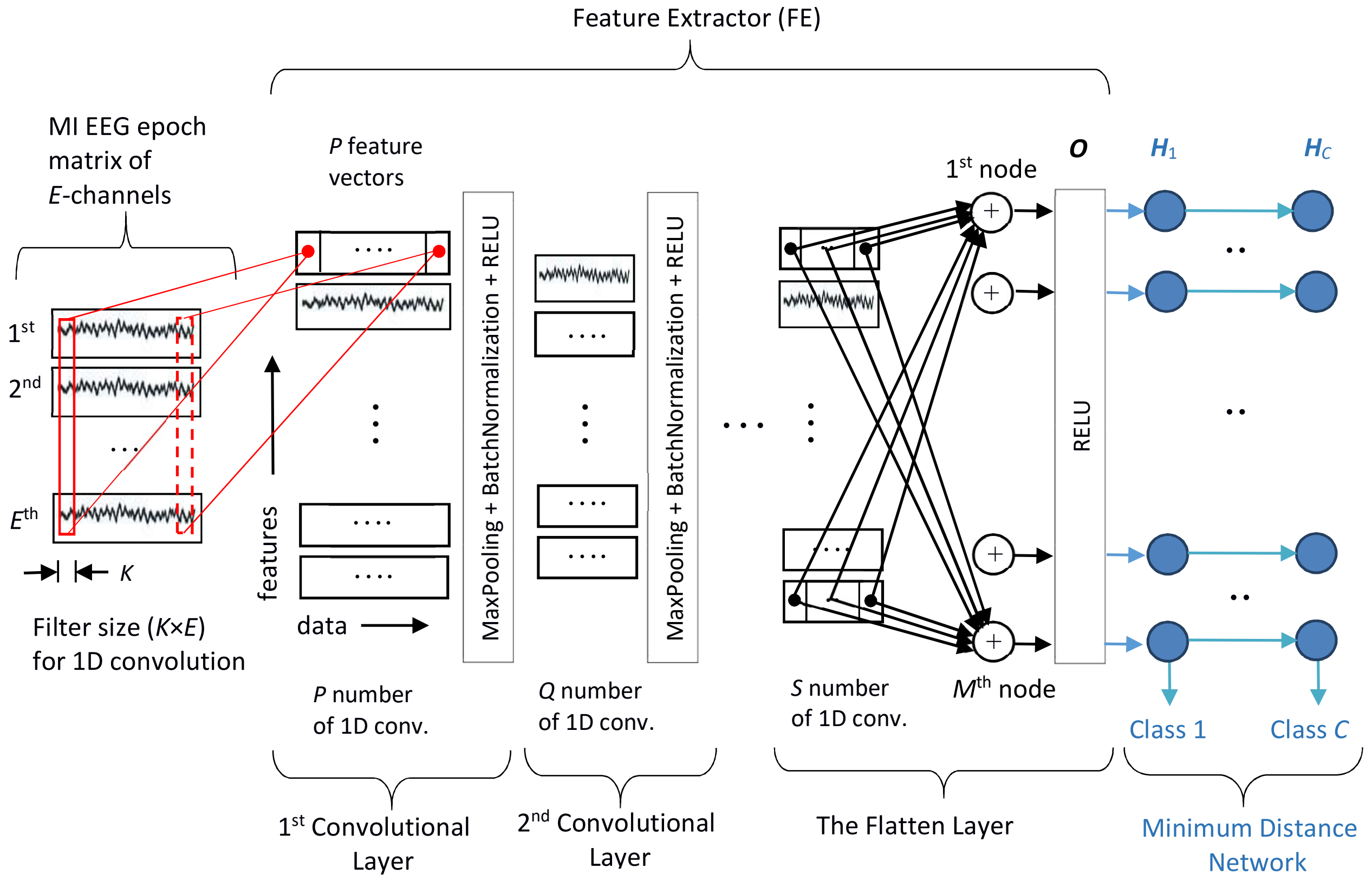

**Fig. 2** The DivFE's structure [33]. $\boldsymbol{O}$ represents the flatten layer. The size of the $\boldsymbol{O}$ vector is equal to the size of Walsh vectors. The FE's structure in our network is the same as the FE of the classical CNN. In this architecture, a minimum distance network (MDN) is proposed instead of the softmax layer. $\boldsymbol{H}_c$ is the $c$th node of the MDN created by using the corresponding row of Walsh matrix. The elements of the $\boldsymbol{H}_c$vector are not affected by the training process.

In this study, we built up a network that aims to maximize the divergence for the feature space so that the CNN can converge it more easily. Maximum divergence for the feature space refers to the strength of the features in intra-class representation and inter-class discrimination. In this respect, a class separability definition is made as given in the following equation:

$$\text{divergence value} = \mathrm{tr}(\mathbf{S}^{-1} \cdot \mathbf{B})$$

$$\mathbf{S} = \mathbf{S}_1 + \cdots + \mathbf{S}_k + \cdots + \mathbf{S}_C \tag{2}$$

In this equation, $C$ is the number of classes, $\mathbf{S}_k$ refers to the covariance matrix of the $k$th class, the trace operation is denoted by tr(.), and $\mathbf{S}$ is the sum of all the covariance matrices. Being the measure of how far apart class centers are from each other, the **B**, is the covariance of the mean vectors of all classes. In the literature, **S** and **B** are also called within-class scatter matrix and between-class scatter matrix, respectively. As seen in the Equation 2, a high divergence value is obtained by minimizing the within-class scatters and maximizing the between-class scatters.

In this respect, ensuring the class centers as distant as possible from each other while the distance between any two classes remains the same across all classes was made successfully by representing each class center by a Walsh vector (a column or a row of the selected Walsh matrix). By associating each row (or column) of the Walsh matrix with a class center, given an input vector, the FE is trained to output the specific row (or column) of the Walsh matrix which represents the class label of that input vector. With this training strategy, selecting the output vectors of the FE as the rows of the Walsh matrix increases the distances between the class centers, thus increases the divergence value which is an indication of efficient features. The **B** matrix is created by using the $\boldsymbol{H}_k$ vectors that are explained in Fig. 2. A symbolic representation of two-dimensional and four-dimensional Walsh matrices is seen in Equation 3. Likewise, their projection to the eight-dimensional space is numerically shown in the same equation by replacing +V and –V values with ones and zeros, respectively.

$$\mathbf{H} = \begin{bmatrix} +V & +V \\ +V & -V \end{bmatrix} \quad \mathbf{H} = \begin{bmatrix} +V & +V & +V & +V \\ +V & -V & +V & -V \\ +V & +V & -V & -V \\ +V & -V & -V & +V \end{bmatrix} \quad \mathbf{H} = \begin{bmatrix} 1 & 1 & 1 & 1 & 1 & 1 & 1 & 1 \\ 1 & 0 & 1 & 0 & 1 & 0 & 1 & 0 \\ 1 & 1 & 0 & 0 & 1 & 1 & 0 & 0 \\ 1 & 0 & 0 & 1 & 1 & 0 & 0 & 1 \\ 1 & 1 & 1 & 1 & 0 & 0 & 0 & 0 \\ 1 & 0 & 1 & 0 & 0 & 1 & 0 & 1 \\ 1 & 1 & 0 & 0 & 0 & 0 & 1 & 1 \\ 1 & 0 & 0 & 1 & 0 & 1 & 1 & 0 \end{bmatrix} \tag{3}$$

In the modified Walsh matrix, it can easily be noticed that the Hamming distance between any two rows or columns is equal to the half of the matrix rank value (rank is equal to the dimension of output vectors). Hence, the value of the MDN nodes for each class label is selected amongst the $\boldsymbol{H}_k$ vectors. Increasing the matrix rank also increases the Hamming distances between the class centers, contributing to the class separability criterion. On the other hand, over increasing the rank should be avoided as it causes prolonged training phases. Therefore, the rank of the Walsh matrix should be selected carefully in order to avoid underfitting and prolonged training time. The training algorithm of the DivFE is demonstrated in Fig. 3.

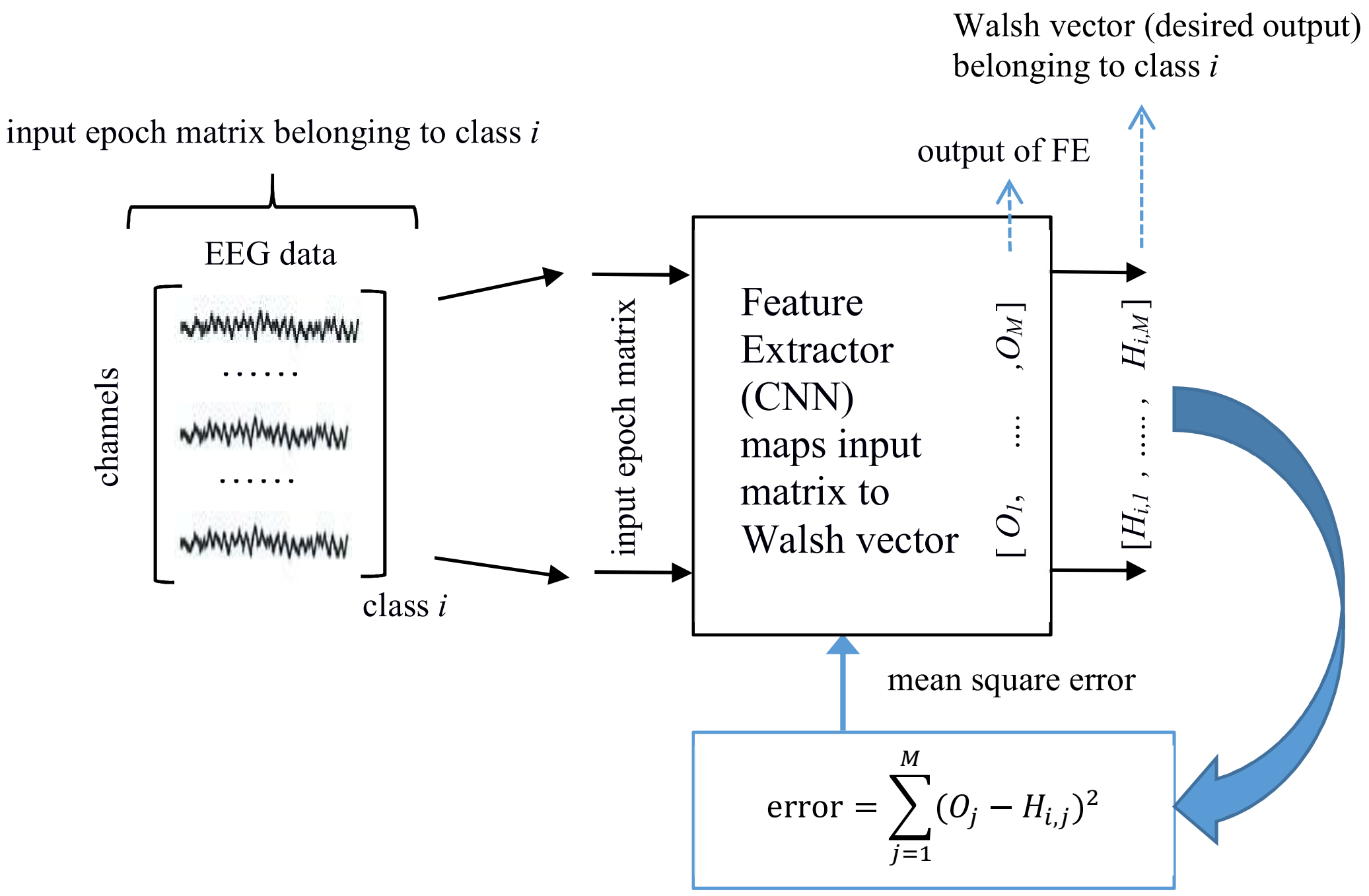

**Fig. 3** DivFE's training procedure. $O_j$ is the *j*th output of the feature extractor, and $H_{i,j}$ is the *j*th element of the Walsh vector associated with the *i*th class. *M* is the rank of the Walsh matrix and is selected as 16 for this study. If we need to represent two different class outputs, the Walsh vectors could be chosen as $\boldsymbol{H}_1$ =[1,0,1,0,1,0,1,0,1,0,1,0,1,0,1,0] and $\boldsymbol{H}_2$ =[1,1,0,0,1,1,0,0,1,1,0,0,1,1,0,0].

### 2.3 Dataset preparation and validation process for the DivFE

Combined MI signals (left hand-right hand, left hand-feet, left hand-tongue, right hand-feet, right hand-tongue, and feet-tongue) are artificially created by using four simple MI signals (left hand, right hand, feet, and tongue) of the BCI Competition III and IV datasets. For instance, in order to generate a combined MI signal corresponding to the tongue-feet, the tongue MI signal is simply added to the feet MI signal as follows:

$$\boldsymbol{EEG}_{C_TF}\{i\}(k,n) = [\boldsymbol{EEG}_{S_T}\{i\}(k,n) + \boldsymbol{EEG}_{S_F}\{i\}(k,n)]/2 \quad (4)$$

where *k* and *n* represent the channels and samples, respectively; $\boldsymbol{EEG}_{C_TF}\{i\}(k)$ represents the *i*th tongue-feet combined MI epoch; $\boldsymbol{EEG}_{S_T}(k)$ and $\boldsymbol{EEG}_{S_F}(k)$ represent the simple tongue and feet MI epochs, respectively. In this way, all the combined MI signals are artificially generated for each subject. The number of combined MI EEG signals is equal to the number of samples of tongue or feet.

The structure of a basic DNN and DivFE, along with the training and validation processes, is described in Algorithm 1. Validation set is made up of 10% of training data. After all the training data is used for once for the training of FE (it is known as one iteration), accuracy and mean squared error loss values are calculated for all data in the validation set. In the validation phase, each input data (epoch) of the validation set is given to the FE (shown in Fig. 2), and FE generates an output vector for each input data. The generated output vector $\boldsymbol{O}$ serves as an input for MDN. The final classification is done by calculating the closest node of MDN to the output vector $\boldsymbol{O}$ which determines the class of the vector. After the assigned Walsh vectors for each input data are compared with the vectors generated by the MDN, mean accuracy and loss values are calculated over all the data in the validation set.

---

**Algorithm 1:** Differences between the training and validation processes, and the structures of a basic DNN and DivFE for two classes [33]

---

**Basic DNN:** Feature Extractor (FE is the convolutional neural network) + Classifier (FCNN)

**1- One iteration of training process for two classes**

**for** each input data in {training set} **do**
  **run** the network (FE + FCNN) and obtain the *network output* $\boldsymbol{O} = [O_1, O_2]$
  **define** *desired output*: $[0, 1]^T$ for the input data of the first class
    $[1, 0]^T$ for the input data of the second class
  **update** weights of network (FE+ FCNN) by calculating $\sum$(*network output* – *desired output*)$^2$
**end for**

**2- Validation/Test process for two classes**

**for** each input data in {validation/test set} **do**
  **run** the network (FE + FCNN) and obtain the *network output* $\boldsymbol{O} = [O_1, O_2]$
  **obtain** the decision "$i$" of basic DNN by calculating the below equation

$$O_i = \max(O_1, O_2)$$

**end for**
**calculate** accuracy by comparing the decision of basic DNN for each input data with the class label of the input

**DivFE** shown in Fig. 1: Feature Extractor (FE is the convolutional neural network) + Classifier (MDN)

**1- One iteration of training process for two classes**

**for** each input data in {training set} **do**
  **run** the network (FE) and obtain the *network output* $\boldsymbol{O}$
  **define** *desired output*: $\boldsymbol{H}_1 = [1,0,1,0,1,0,1,0,1,0,1,0,1,0,1,0]^T$ for the input data of the first class
    $\boldsymbol{H}_2 = [1,1,0,0,1,1,0,0,1,1,0,0,1,1,0,0]^T$ for the input data of the second class
    Except $\boldsymbol{H}_i = [1,1,1,1,1,1,1,1,1,1,1,1,1,1,1,1]^T$
  **update** weights of network (FE) by using $\sum$(*network output* – *desired output*)$^2$
**end for**

**2- Validation/Test process for two classes**

**for** each input data in {validation/test set} **do**
  **run** the network (FE) and obtain the *network output* $\boldsymbol{O} = [O_1, \dots, O_{16}]$
  **obtain** the decision "$i$" of DivFE by calculating the below equation for $\boldsymbol{H}_1$ and $\boldsymbol{H}_2$

MDN: $\boldsymbol{D}_k = \sum_{j=1}^{M} (O_j - H_{k,j})^2 \qquad \boldsymbol{D}_i = \min_k(\boldsymbol{D}_k)$

**end for**
**calculate** accuracy by comparing decision of DivFE for each input data with the true class label of the input

---

To avoid under and over fitting related issues, mean squared error (MSE) and accuracy values for both the training and validation datasets are examined. Relevant to the literature, the hyper-parameters of the FE network, *i.e.* filter length, number of layers, number of features in each layer, *etc.*, are determined using trial and error method. The training for longer iterations is started once the coarse structure of the FE network is finalized. Validation dataset's loss values and number of iterations are the parameters that are considered for the termination of training. For $N$ randomly generated training and test sets, the FE is trained $N$ times and the average of $N$ accuracies for the test sets is calculated in order to demonstrate the statistical significance of the results.

## 3. COMPUTER SIMULATIONS

Data analyses were done using Python on an Ubuntu Linux based workstation. The CPU had 32 cores which are clocked at 2.7 GHz. In addition, the workstation was equipped with a GTX2080 Ti graphics card. The benchmark datasets, namely BCI Competitions III (2005) and IV (2008) databases [1, 2], are utilized for testing the proposed modifications in the CNN structure, and training this structur for MI signal classification.

Table 1 provides a summary of the recent studies performed for the classification of MI EEG signals by using the DNNs. The four rows at the bottom of the table show the classification results of our previous study [33]. Table 2 provides a summary of the recent studies performed for the classification of MI EEG signals by using the traditional neural networks. It is observed that success rates obtained with DNNs are higher than those obtained with traditional neural networks. In addition, one of the advantages of DNNs is that there is no need to spend an extra effort to determine the features; the CNN automatically extracts the features from the dataset during the training.

### 3.1 ERD/ERS ratio analyses for simple and combined MI signals

In this section, ERD/ERS ratios will be plotted and studied for the combined MI signals artificially produced using the simple MI signals. Especially the left hand, right hand, and both hands MI signals will be analyzed for C3 and C4 electrodes and the results will be compared with the graphs in the study [36]. In this context, the advantages and disadvantages of generating combined MI signals using simple MI signals will be discussed. Then, the classification results for simple and combined MIs obtained by using the BCI Competitions III and IV datasets will be given.

In BCI Competition IV dataset [2], 22 electrodes were used to record the EEGs; the montage is shown in Fig. 4. The signals were sampled at 250 Hz, and the data set is made up of the EEG recordings from nine subjects. The paradigm is cue-based and contains the MI tasks of the left hand, right hand, both feet, and tongue. For each subject, two sessions that are comprised of six runs were recorded on different days. One run consists of 48 trials (12 for each class), yielding a total of 288 trials per session. The parameters used in the dataset collected in BCI Competition IV-IIa (such as electrode number, sampling frequency) are almost the same as the parameters used in the study [36].

**Table 1.** The classification results of the MI EEG signals with two and four classes by using DNNs.

| studies | database-dataset | number of classes | mean accuracy % | transformation | number of subjects |
|---|---|---|---|---|---|
| Yang's study[3] | IV-IIa | 4 | 69.27 | Filter + CSP | 9 |
| Sakhavi's study [4] | IV-IIa | 4 | 70.6 | Filter + CSP | 9 |
| Lu's study [5] | IV-IIb | 2 | 84 | FFT | 9 |
| Sakhavi's study [6] | IV-IIa | 4 | 74.46 | Filter + CSP | 9 |
| Abbas's study[7] | IV-IIa | 4 | 70.7 | Filter + CSP | 9 |
| Wu's study[8] | IV-IIb | 2 | 80.6 | Filter bank | 9 |
| Dai's study[9] | IV-IIb | 2 | 78.2 | STFT | 9 |
| Tabar's study[10] | IV-IIb | 2 | 77.6 | STFT | 9 |
| Tang's study[11] | III-IIIa | 4 | 91.9 | CSP | 3 |
| Chaudhary's study[12] | III-IVa | 2 | **99.3** | CWT | 5 |
| Zhao's study[13] | IV-IIa | 4 | 75 | 3D-EEG | 9 |
| Zhang's study [14] | IV-IIb | 2 | 82 | CSP+bispectrum | 9 |
| Deng's study [17] | III-IIIa | 4 | 85.3 | FBCSP | 3 |
| | IV-IIa | 4 | 78.9 | FBCSP | 9 |
| Olivas-Padilla's study [18] | IV-IIa | 4 | 78.4<br>for monolithic network | DFBCSP | 9 |
| Liu's study [19] | IV-IIa | 4 | 76.86 | CSP | 9 |
| Dokur's study [33] | III-IVa | 2 | 96.2<br>98.5 | no transformation<br>CSP | 5 |
| Dokur's study [33] | III-IIIa | 4 | **96.5**<br>95.8 | no transformation<br>CSP | 3 |
| Dokur's study [33] | IV-IIa | 4 | **79.3**<br>79.1 | no transformation<br>CSP | 9 |
| Dokur's study [33] | IV-IIb | 2 | **88.6**<br>85.1 | no transformation<br>CSP | 9 |

CSP : Common spatial patterns
CWT : Continuous wavelet transform
DFBCSP: Discriminative Filter bank common spatial patterns
FBCSP: Filter bank common spatial patterns
FFT : Fast Fourier transform
STFT : Short-time Fourier transform

The modulation of ERD is known to be specific at 8–12 Hz range. Subsequently, the oscillatory power of the grand average across trials was estimated separately for the simple and combined MIs by squaring its samples and smoothing the resulting signal by using a 0.32 second sliding window with a 4 ms shifting step. The signals in the graphs are normalized by dividing the instantaneous powers in all channels by the average of the instantaneous power at the 2.5th second. Fig. 5(a) shows, respectively, the grand average of the resulting oscillatory power for the left hand, right hand, and both hands in electrodes C3 and C4.

**Table 2**. The classification results of MI EEG signals with the traditional neural networks.

| Methods | accuracy % | number of classes | transformation + feature extraction | classifier | dataset |
|---|---|---|---|---|---|
| Wang’s method [20] | 77.2 | 4 | CSP + variance | MLP | BCI III-IIIa |
| Aljalal's method [21] | 80.2 | 2 | Wavelet + statistical, entropy, energy features | MLP | BCI III-IVa |
| Mirnaziri's method [22] | 61.7 | 4 | CSP + variance | MLP | BCI IV-IIa |
| Silva's method [23] | 67.8 | 2 | Linear Predictive Coding | MLP | BCI IV-IIb |
| Alansari’s method [24] | 83.8 | 2 | Wavelet | SVM | BCI IV-IIb |
| Behri’s method [25] | 89.4<br>94.5 | 2 | Wavelet | SVM<br>K-NN | BCI III-IVa |
| Zhang’s method [26] | 84 | 2 | CSP + variance | SVM | BCI III-IVa |
| Li’s method [27] | 68.6 | 3 | FBCSP+ variance | SVM | BCI IV-IIa |
| Wang’s method [28] | 81.2 | 2 | FD-CSP + variance | SVM | BCI IV-IIb |
| Mishuhina’s method [29] | 89.8 | 4 | RCSP – FWR | LDA | BCI III-IIIa |
| Molla’s method [30] | 92.2<br>91.3 | 2 | CSP + subband features | SVM<br>LDA | BCI III-IVa |

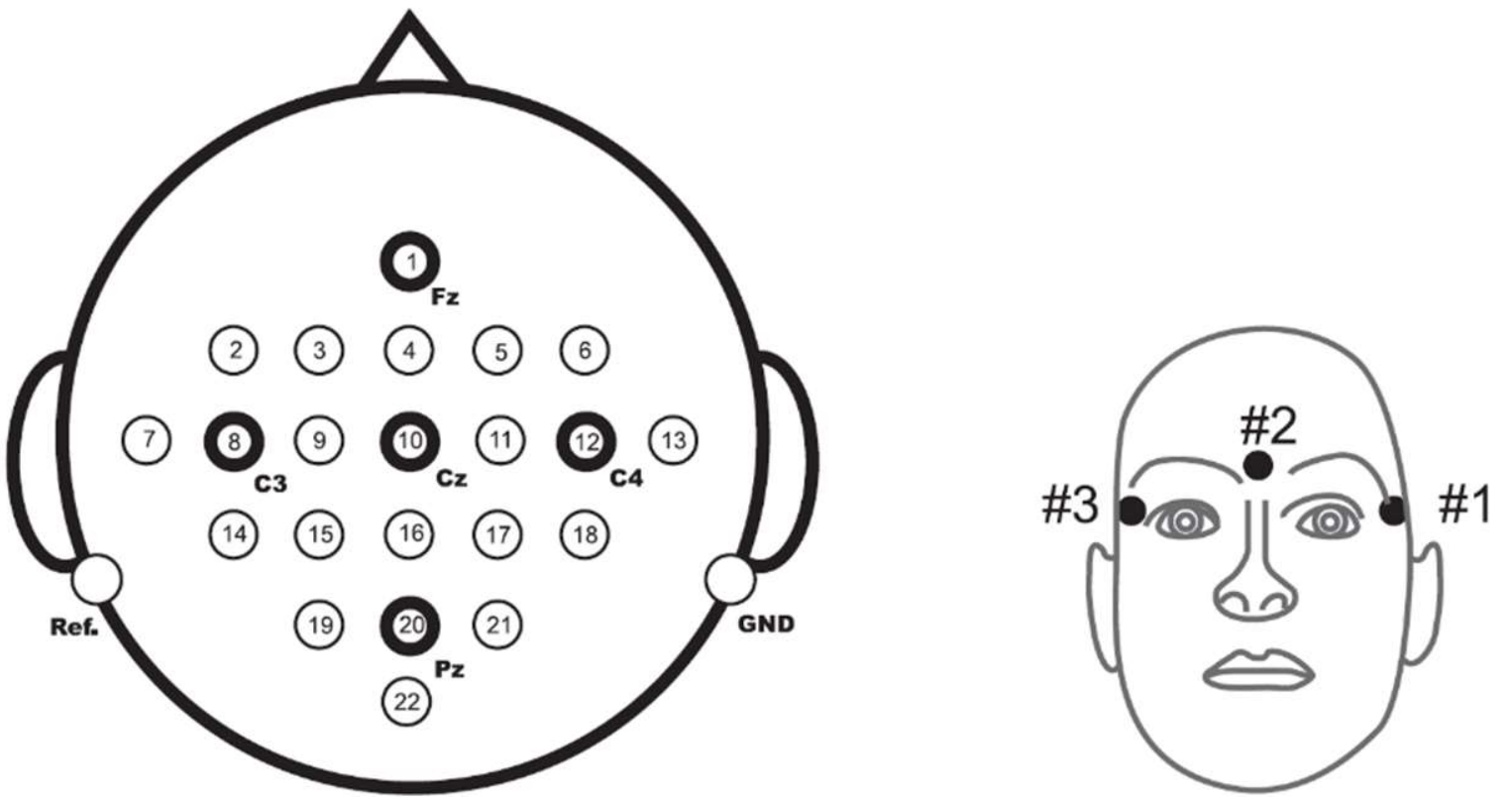

Figure 4. The montage of the BCI Competition IV dataset [2].

As can be seen in Fig. 5, similar results are obtained compared with the graphs shown in Figs. 3 and 4 of the study [36] for the MIs of the left hand, right hand, and both hands. This similarity shows that the artificially produced combined MI signal has the same characteristics with the combined MI signal acquired directly from the subjects in the study [36]. Therefore, this artificially produced MI signal can be used as if it were a real signal acquired from a subject. The artificially produced MI signal actually shows the ideal situation. Some subjects have difficulty generating even a single MI signal. Therefore, it will be even more difficult to generate two MIs at the same time for some subjects. The approach to artificially create MI signals described in this paper can provide feedback for these subjects during experiments. With this feedback, the subject can learn to generate two motor imageries (combined MI signal) at the same time.

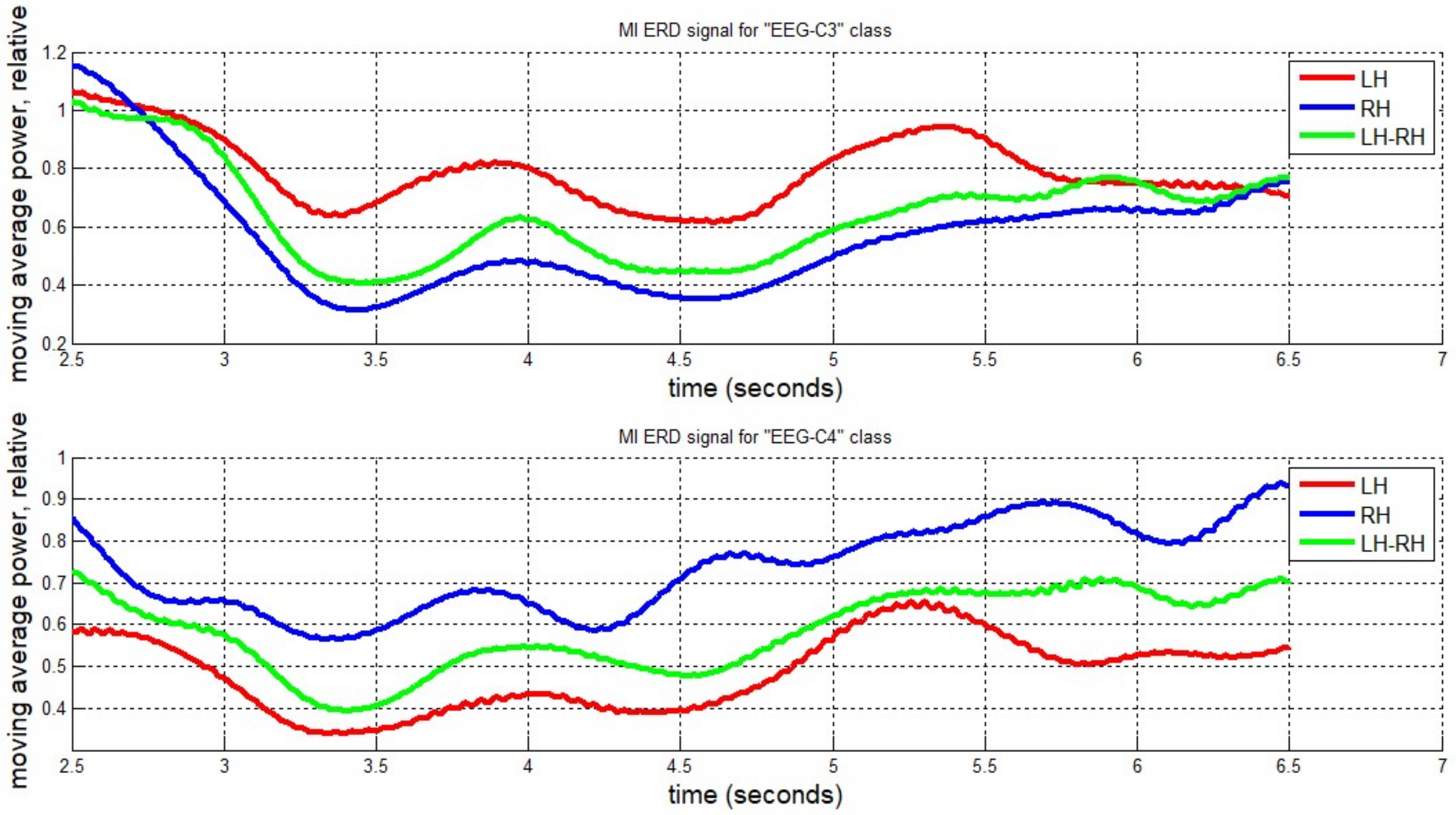

(a)

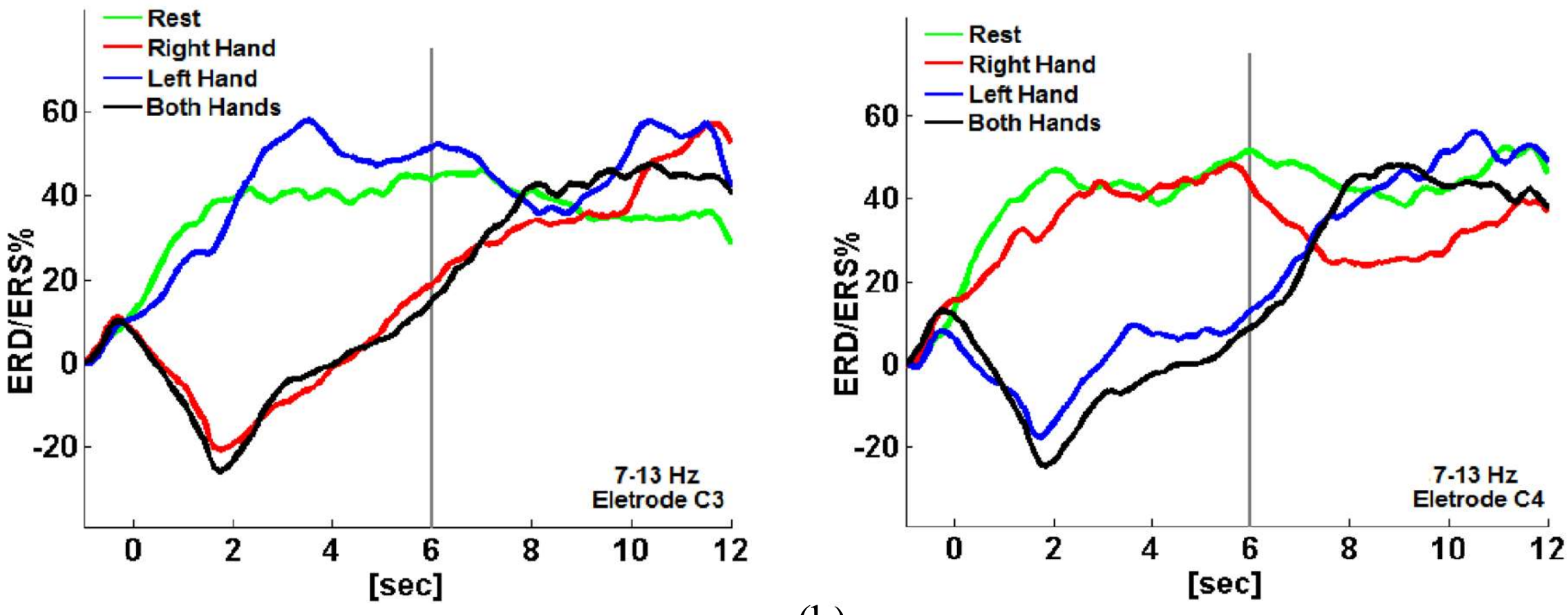

(b)

Figure 5(a) ERD/ERS ratios of simple and combined MI tasks according to the C3 and C4 electrodes, and (b) similar results obtained for simple and combined MI tasks in the study [36].

Fig. 6 shows relative average power at the electrodes C3, CZ and C4 according to the simple (left hand, right hand, and feet) and combined (left hand-right hand, left hand-feet, and right hand-feet) MI tasks. In the study [36–41], researchers have studied on the three simple MI tasks and the combined MI tasks of them. In this study, we have dealt with four simple MI tasks and their combinations. Fig. 7 shows relative average power at the electrodes C4, C3, CZ and C13 according to the simple (left hand-LH, right hand RH, feet F and tongue T) and combined (left hand-right hand LH-RH, left hand-feet LH-F, left hand-tongue LH-T, right hand-feet RH-F, right hand-tongue RH-T and feet-tongue F-T) MI tasks.

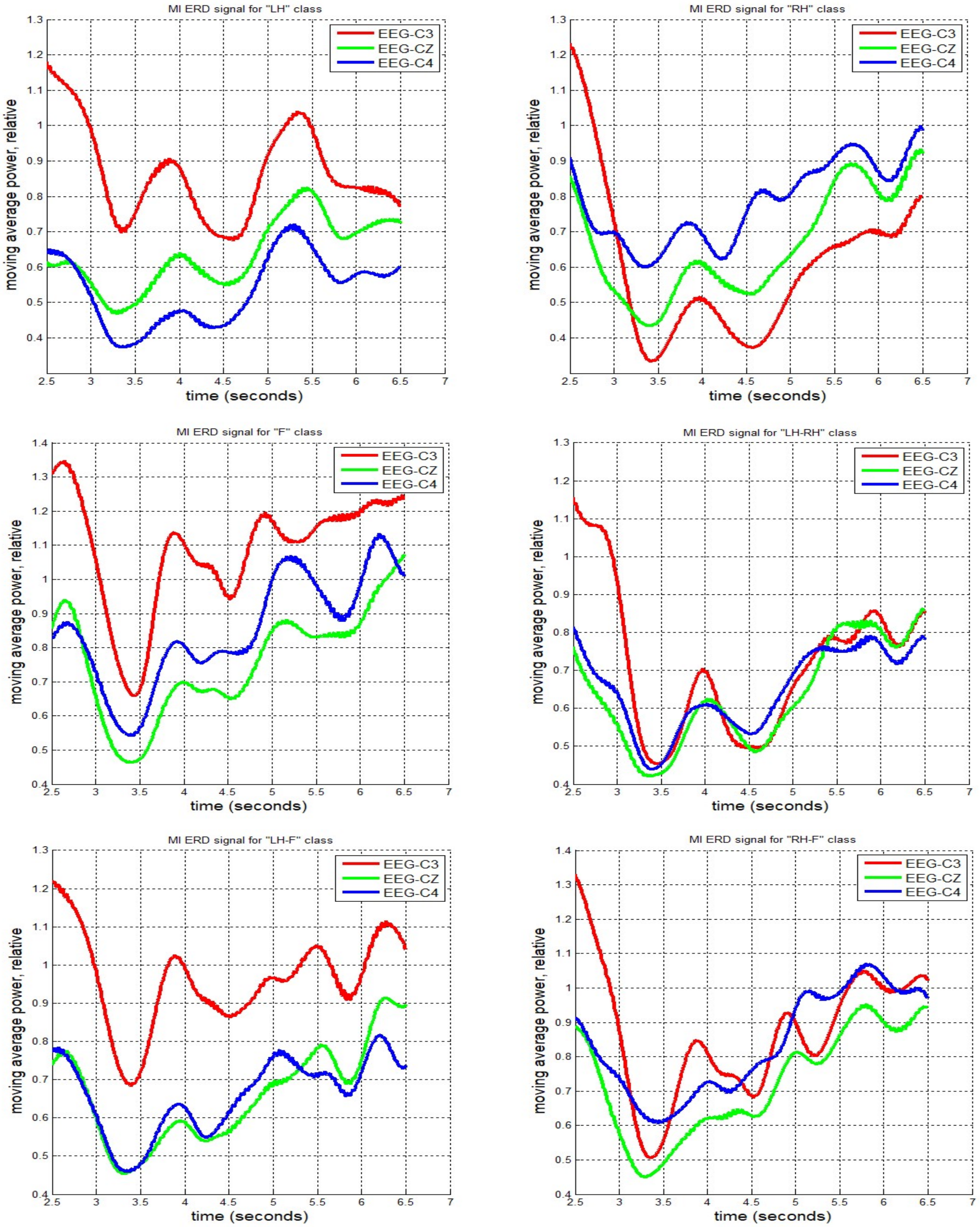

**Fig. 6.** Relative average power at the electrodes C3, CZ and C4 (in Fig. 4) according to the simple (left hand, right hand, and feet) and combined (left hand-right hand, left hand-feet, and right hand-feet) MI tasks for the subject S3 in BCI dataset II-a.

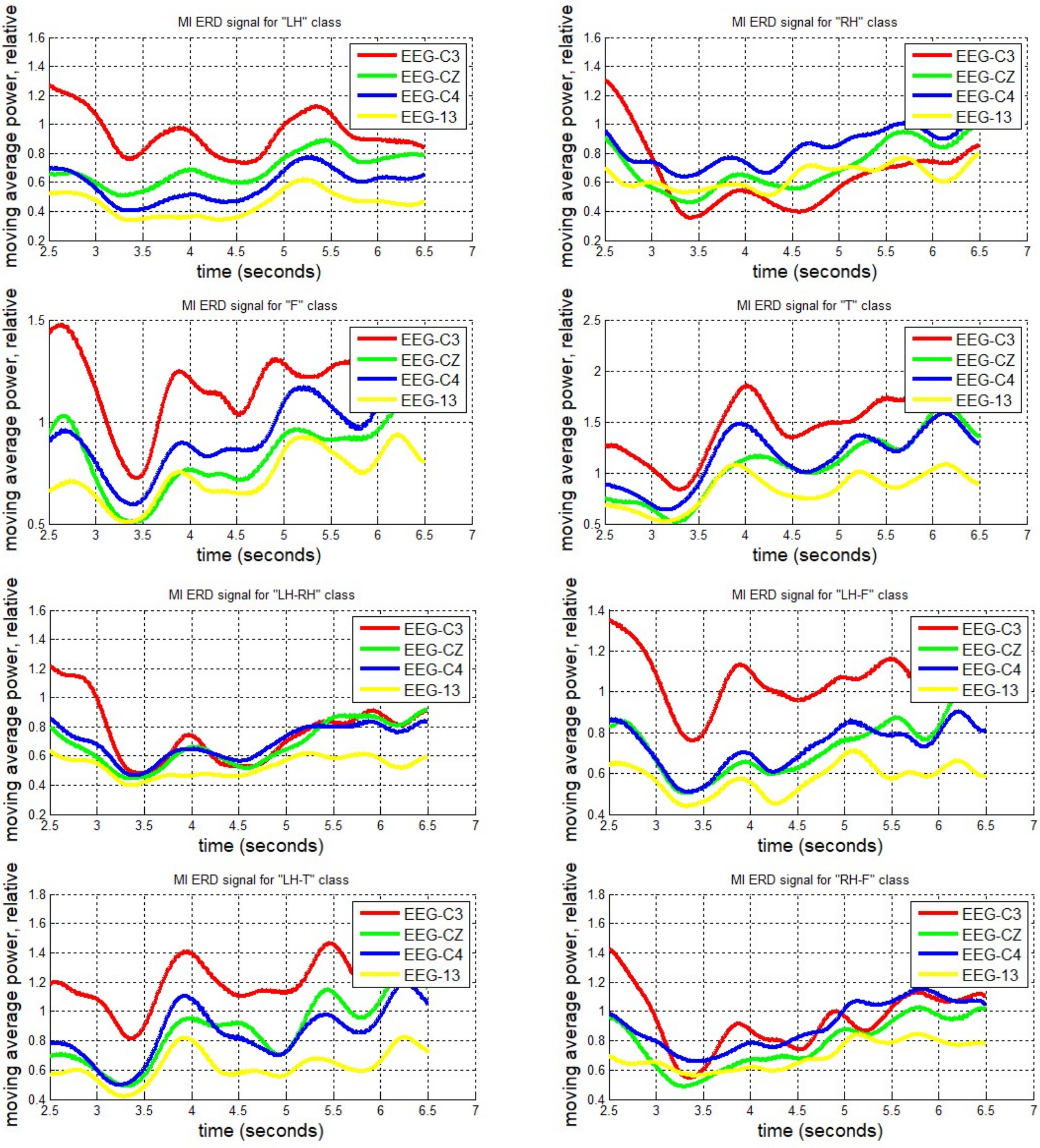

**Fig. 7.(a)** Relative average power at the electrodes C3, CZ, C4 and C13 (in Fig. 4) according to the simple and combined MI tasks for the subject S3 in BCI dataset II-a.

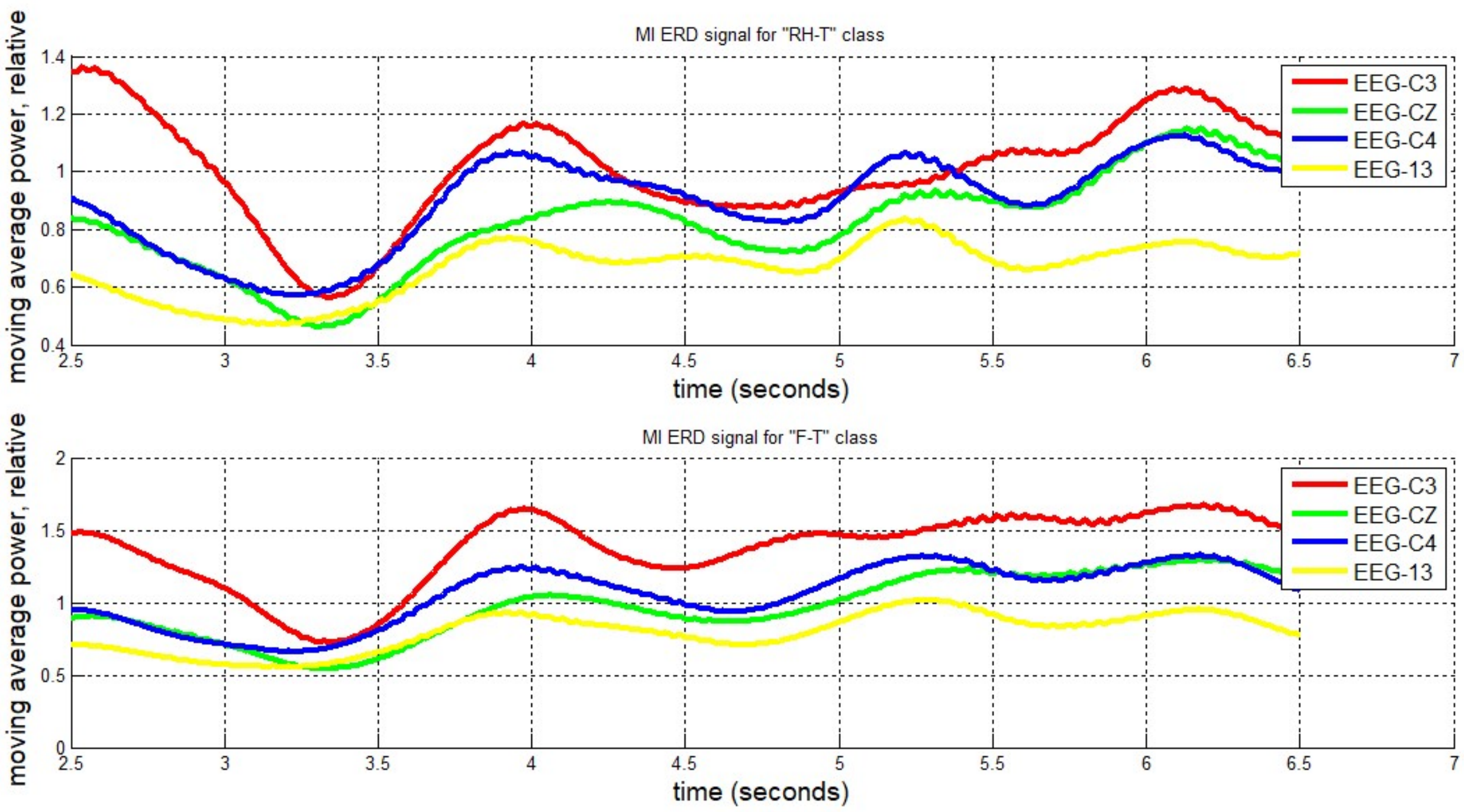

**Fig. 7.** (continued) **(b)** Relative average power at the electrodes C3, CZ, C4 and C13 (in Fig. 4) according to the simple and combined MI tasks for the subject S3 in BCI dataset II-a.

Fig. 7. shows that the simple and combined MI EEG signals can easily be classified by using the DNNs. It is observed that for more number of MI tasks, it will be harder to distinguish the classes from each other. Moreover, more than four electrodes were used to classify the MI EEG signals.

### 3.2 Validation on the dataset of the simple and combined MI EEG signals acquired in real time

In the dataset [40], subjects were seated in a comfortable chair with the arms at their sides in front of a screen showing the task cue to be performed, which consisted of one of the eight mental states were generated with all the combinations of the right hand, left hand, and both feet together, *i.e.*, rest, left hand, feet, left hand-feet, right hand, both hands, right hand-feet, and both hands-feet. The whole session consisted of four runs, containing 10 trials per task for a total of 40 trials per class (320 trials considering the eight classes). Each trial was randomly presented and lasted for 12 seconds, starting at time '0' with a cross at the center of each panel and an overlaid arrow indicating for the next 6 seconds the motor task to be performed. EEG signals were recorded at 256 Hz using a commercial amplifier. Both the signal acquisition and the stimulation process were implemented on the OpenViBE platform. The EEG cap was fitted with 26 electrodes, namely, Fp1, Fpz, Fp2, Fz, FC5, FC3, FC1, FCz, FC2, FC4, FC6, C5, C3, C1, Cz, C2, C4, C6, CP5, CP3, CP1, CPz, CP2, CP4, CP6, and Pz, re-referenced with respect to the common average reference across all channels and located over the extended international 10–20 system positions to cover the primary sensorimotor cortex. Signals were band-pass filtered within the frequency range (8–30 Hz) using a fifth-order Butterworth filter [40].

It is known that MI EEG signals are highly dependent on the subjects. In this study, we did not deal with person-independent MI EEG signals. Therefore, each subject's own data was used for validation. In other words, DivFE trained with data collected from one subject was not used to classify the data of other subjects. In the dataset [40], there are eight classes: Left hand (LH), right hand (RH), feet (F), left hand-feet (LHF), right hand-feet (RHF), both hands (BH), both hands-feet (HsF) and rest (R). Left hand-feet, right hand-feet, and left hand-right hand combined MI EEGs are generated artificially by using the simple feet, left hand, and right hand MI EEG data of [40] according to the calculation given in Equation 4. All of the artificially generated MI EEG signals (LHF, RHF and BH) are included into the training set; and all of the real combined MI EEG signals (LHF, RHF and BH) are included into the test set. 80% of the LH, RH, F, R, and HsF MI EEG signals in [40] is used in the training set, and the remaining 20% of the MI EEGs in [40] is reserved for the test set.

In order to compare the results obtained in our study with the results in [40], one versus all (OVA) method is preferred in this study. Therefore, eight different DivFE networks are used to determine the eight classes left hand versus others (LH-O), right hand versus others (RH-O), feet versus others (F-O), BH versus others (BH-O), LHF versus others (LHF-O), RHF versus others (RHF-O), HsF versus others (HsF-O), and R versus others (R-O). Moreover, due to the use of OVA method, the number of data in the training set is not balanced. The augmentation process is used to eliminate the imbalance within the training set. Tables 3(a) and (b) show the number of epochs related to the classes in the training and test sets for LH-O and BH-O classes, respectively; there are 40 trials per class in the dataset [40]. In the LH-O experiment, the O class (blue colored) consists of RH, F, BH, LHF, RHF, HsF and R classes. In this section, all the tables show the averages of 30 experiments with randomly partitioned training and test sets.

**Table 3a.** The training and test sets for the LH-O (left hand versus others) class.

| | LH | RH | F | BH | LHF | RHF | HsF | R |
|---|---|---|---|---|---|---|---|---|
| all epochs | 8r + 32r | 40r | 40r | 40a+40r | 40a+40r | 40a+40r | 40r | 40r |
| training set | 248u | 32r | 32r | 40a | 40a | 40a | 32r | 32r |
| test set | 8r | 8r | 8r | 40r | 40r | 40r | 8r | 8r |

40r: 40 real epochs
40a: 40 artificial epochs
248u: augmented epochs obtained from 32 real epochs in the training set

Table 4 shows the classification performances obtained for the subjects of the dataset in [40]. In Table 4, successes are defined as the averages of accuracies obtained for all classes; it can be observed that high success rates have been achieved for each subject. Table 5 shows the confusion matrices obtained for the subject S2. While obtaining the results in Table 4, it is aimed to keep the sensitivities high. In this respect, Table 5 reveals our concern about achieving high sensitivity values.

**Table 3b.** The training and test sets for the BH-O (both hands versus others) class.

| | LH | RH | F | BH | LHF | RHF | HsF | R |
|---|---|---|---|---|---|---|---|---|
| all epochs | 40r | 40r | 40r | 240a+40r | 40a+40r | 40a+40r | 40r | 40r |
| training set | 32r | 32r | 32r | 240a | 40a | 40a | 32r | 32r |
| test set | 8r | 8r | 8r | 40r | 40r | 40r | 8r | 8r |

40r: 40 real epochs
40a: 40 artificial epochs
240a: 240 artificial epochs

**Table 4.** The classification accuracies and (Kappa) for the subjects of the datasets in [40].

| classes | % accuracies and (Kappa) values for the subjects S1, S2, … , S7 | | | | | | | |
|---|---|---|---|---|---|---|---|---|
| | S1 | S2 | S3 | S4 | S5 | S6 | S7 | % mean accuracy |
| LF-O | 69.2 | 86.3 | 80.0 | 78.1 | 78.1 | 78.8 | 86.3 | |
| RH-O | 76.7 | 86.3 | 81.9 | 73.8 | 79.4 | 83.8 | 76.9 | |
| F-O | 75.8 | 81.9 | 65.0 | 76.9 | 75.6 | 89.4 | 77.5 | |
| BH-O | 71.7 | 80.0 | 68.1 | 59.4 | 74.3 | 68.1 | 70.0 | |
| LHF-O | 61.7 | 74.4 | 78.1 | 68.1 | 71.9 | 60.0 | 75.6 | |
| RHF-O | 72.5 | 76.3 | 72.5 | 68.1 | 63.8 | 70.6 | 63.1 | |
| HsF-O | 81.7 | 81.3 | 62.5 | 68.8 | 76.3 | 73.8 | 80.0 | |
| R-O | 83.3 | 97.5 | 92.5 | 81.3 | 96.3 | 90.0 | 93.0 | |
| accuracy (Kappa) in this study | 74.1 (0.482) | 83.0 (0.66) | 75.1 (0.502) | 71.8 (0.346) | 76.9 (0.538) | 76.8 (0.536) | 77.8 (0.556) | 76.5 |
| accuracy (Kappa) by OVA in [40] | 33.7 | 67.5 (0.35) | 46.5 | 31.8 | 35.9 | 49.3 | 58.7 (0.174) | 46.2 |

A DivFE of similar structure was used for each subject. The input signal size is selected as 1536×1 (256×6 seconds). In the DivFE architecture, the feature extractor has eleven convolutional layers and a single dense layer. The size of the filters is 45 in each layer, and there are 70 feature maps in each convolution layer. In these experiments, max-pooling process is not used. The single dense layer has sixteen output nodes due to using sixteen-dimensional Walsh vectors for the training; so, it is composed of 70×1536×16 weights. The number of weights of the DivFE is calculated as follows:

$$1\times45\times70 + [70\times45\times70]\times10 + 70\times1536\times16 = 3928470$$

**Table 5**. The confusion matrices for the subject S2

| 86.3% accuracy | LH | O |
|---|---|---|
| LH | 6 | 2 |
| O | 20 | 132 |

| 86.3% accuracy | RH | O |
|---|---|---|
| RH | 7 | 1 |
| O | 21 | 131 |

| 81.9% accuracy | F | O |
|---|---|---|
| F | 7 | 1 |
| O | 28 | 124 |

| 80% accuracy | BH | O |
|---|---|---|
| BH | 23 | 17 |
| O | 15 | 105 |

| 74.4% accuracy | LHF | O |
|---|---|---|
| LHF | 27 | 13 |
| O | 28 | 92 |

| 76.3% accuracy | RHF | O |
|---|---|---|
| RHF | 28 | 12 |
| O | 26 | 94 |

| 81.3% accuracy | HsF | O |
|---|---|---|
| HsF | 5 | 3 |
| O | 27 | 125 |

| 97.5% accuracy | R | O |
|---|---|---|
| R | 8 | 0 |
| O | 4 | 148 |

The paired t-test is applied to the results to show that the achievements in Table 4 are not random. For this purpose, two sets are created with the sensitivities of the results obtained from the seven subjects for randomly generated epochs (RGE) and artificially generated epochs (AGE) (by using Equation 4), respectively. In these experiments, only BH-O, LHF-O and RHF-O sensitivity values obtained for each subject are used. Therefore, the first set is created with a total of 21 data (three data from each subject) by using the sensitivities of results in Table 4. In order to form the second set, a different experiment is created. In this context, epochs with random values are replaced with the artificial epochs (240a and 40a) in the training set while the test set remains unchanged as shown in Table 3b, and DivFEs are trained individually for each subject. The sensitivities corresponding to the BH, LHF and RHF classes for seven subjects are used to form the second set. The violin plot provided in Fig. 3 allows for a visual investigation of the effect of artificially generated epochs on the accuracy of the classification results. The paired t-test further validated that, the obtained accuracy results with artificially generated epochs are statistically significantly higher (p-val: 1.82e–18) than the results obtained with randomly generated epochs.

### 3.3 Classification of the simple and combined MI EEG signals by using DNNs

Table 6 shows the classification results (sensitivities and mean accuracy) of the DivFE for the ten-class MI EEG signals generated from BCI IV-IIa [2]. Table 7 presents the FE structures specifically designed for each subject to generate the classification results in Table 6. Max-pooling, batch normalization and ReLU were used at the output of each convolution layer. Tables 6 and 7 show that high classification performances are achieved by using the DivFEs with a small number of weights for dataset BCI IV-IIa. Moreover, in Table 7, the structures of the networks for the nine subjects are similar with each other. This similarity indicates that the form of the structures is independent of the subjects. Similarity of the structures can be considered as an advantage. Thus, the coarse structure can be determined more easily.

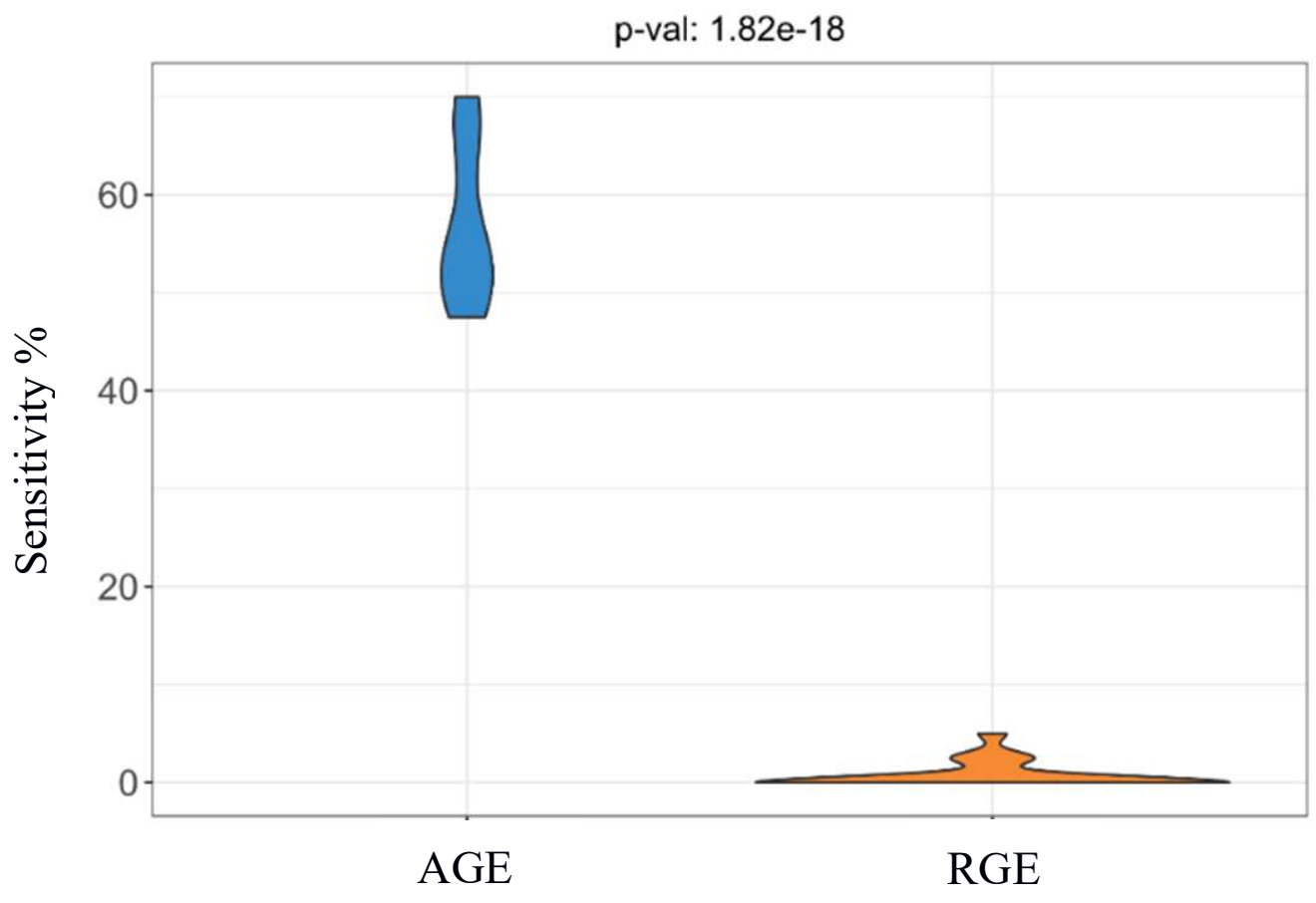

Figure 8. The paired t-test to demonstrate the effect of artificially generated epochs. AGE represents artificially generated epochs and RGE represents randomly generated epochs.

**Table 6.** The classification results of the DivFE for ten-class MI EEGs generated from BCI IV-IIa.

| | % sensitivities obtained for each subject | | | | | | | | | |
|---|---|---|---|---|---|---|---|---|---|---|
| | **S1** | **S2** | **S3** | **S4** | **S5** | **S6** | **S7** | **S8** | **S9** | % mean accuracy (Kappa) |
| class LH | 100 | 92.6 | 97.7 | 91.7 | 100 | 90.5 | 88.9 | 92.3 | 91.2 | |
| class RH | 88.9 | 44.4 | 82.7 | 75 | 69.2 | 81 | 85.2 | 57.7 | 70.8 | |
| class F | 81.5 | 70.4 | 60.2 | 66.7 | 69.2 | 47.6 | 70.4 | 80.8 | 66.7 | |
| class T | 63 | 81.5 | 67.7 | 70.8 | 76.9 | 76.1 | 66.7 | 42.3 | 62.5 | |
| class LH-RH | 88.9 | 88.9 | 94 | 95.8 | 96.2 | 100 | 88.9 | 96.2 | 100 | |
| class LH-F | 59.3 | 70.4 | 52.6 | 79.2 | 69.2 | 66.7 | 85.2 | 73.1 | 62.5 | |
| class LH-T | 44.4 | 63 | 71.4 | 66.7 | 76.9 | 81 | 74.1 | 69.2 | 54.2 | |
| class RH-F | 85.2 | 100 | 86.5 | 95.8 | 88.5 | 95.2 | 92.6 | 96.2 | 100 | |
| class RH-T | 66.7 | 66.7 | 67.7 | 37.5 | 84.6 | 66.7 | 74.1 | 50 | 75 | |
| class F-T | 92.6 | 92.6 | 97.7 | 91.7 | 100 | 100 | 85.2 | 100 | 100 | |
| % accuracy | 77 | 77.1 | 77.8 | 77.1 | 83 | 80.5 | 81.1 | 75.8 | 78.3 | 78.6 (0.715) |

**LH:** left hand
**RH:** right hand
**F:** feet
**T:** tongue

**LH-RH:** left hand-right hand
**LH-F:** left hand-feet
**LH-T:** left hand-tongue

**RH-F:** right hand-feet
**RH-T:** right hand-tongue
**F-T:** feet-tongue

Table 8 shows the classification results (sensitivities and mean accuracy) of the DivFE for the ten-class MI EEG signals created by using the BCI III-IIIa [1] dataset. Table 9 presents the FE structures specifically designed for each subject to generate the classification results in Table 8. Max-pooling, batch normalization and ReLU were used at the output of each convolution layer. Tables 8 and 9 show that high classification performances are achieved by using the DivFEs with a small number of weights for the dataset BCI III-IIIa. Moreover, in Table 9, the structures of the networks for the three subjects are similar with each other.

**Table 7** The FE structures which generate the classification results in Table 6.

| FE Layers | #input nodes, filter size, #feature planes in FE layers | | | | | | | | |
|---|---|---|---|---|---|---|---|---|---|
| | **S1** | **S2** | **S3** | **S4** | **S5** | **S6** | **S7** | **S8** | **S9** |
| **Layer 1** | 22,15,70 | 22,15,70 | 22,15,50 | 22,15,70 | 22,15,70 | 22,15,70 | 22,15,70 | 22,15,50 | 22,15,40 |
| **Layer 2** | 70,13,70 | 70,13,70 | 50,13,50 | 70,13,70 | 70,13,70 | 70,13,70 | 70,13,70 | 50,13,50 | 40,13,40 |
| **Layer 3** | 70,11,70 | 70,11,70 | 50,11,50 | 70,11,70 | 70,11,70 | 70,11,70 | 70,11,70 | 50,11,50 | 40,11,40 |
| **Layer 4** | 70,9,70 | 70,9,70 | 50,9,50 | 70,9,70 | 70,9,70 | 70,9,70 | 70,9,70 | 50,9,50 | 40,9,40 |
| **Layer 5** | 70,7,70 | 70,7,70 | 50,7,50 | 70,7,70 | 70,7,70 | 70,7,70 | 70,7,70 | 50,7,50 | 40,7,40 |
| **Layer 6** | 70,5,70 | 70,5,70 | 50,5,50 | 70,5,70 | 70,5,70 | 70,5,70 | 70,5,70 | 50,5,50 | 40,5,40 |
| **Layer 7** | 70,3,70 | 70,3,70 | 50,3,50 | 70,3,70 | 70,3,70 | 70,3,70 | 70,3,70 | 50,3,50 | 40,3,40 |
| **FL** | 70,4,16 | 70,4,16 | 50,4,16 | 70,4,16 | 70,4,16 | 70,4,16 | 70,4,16 | 50,4,16 | 40,4,16 |

**FL:** flatten layer

**Table 8** The classification results of the DivFE for ten-class MI EEG signals produced from BCI III-IIIa.

| | % sensitivities obtained for each subject | | | |
|---|---|---|---|---|
| | **k3b** | **k6b** | **l1b** | |
| class LH | 87.5 | 100 | 100 | |
| class RH | 67.3 | 75 | 87.5 | |
| class F | 87.5 | 50 | 100 | |
| class T | 74 | 75 | 62.5 | |
| class LH-RH | 87.5 | 100 | 100 | |
| class LH-F | 74 | 75 | 87.5 | |
| class LH-T | 67.3 | 87.5 | 50 | |
| class RH-F | 100 | 100 | 100 | |
| class RH-T | 87.5 | 87.5 | 100 | % mean accuracy (Kappa) |
| class F-T | 80.8 | 100 | 100 | |
| % accuracy | 81.3 | 85 | 88.8 | 85 (0.8) |

**Table 9** The FE structures which generate the classification results in Table 8.

| FE layers | #input nodes, filter size, #feature planes in FE layers | | |
|---|---|---|---|
| | **k3b** | **k6b** | **l1b** |
| **Layer 1** | 43,15,70 | 43,15,70 | 43,15,50 |
| **Layer 2** | 70,13,70 | 70,13,70 | 50,13,50 |
| **Layer 3** | 70,11,70 | 70,11,70 | 50,11,50 |
| **Layer 4** | 70,9,70 | 70,9,70 | 50,9,50 |
| **Layer 5** | 70,7,70 | 70,7,70 | 50,7,50 |
| **Layer 6** | 70,5,70 | 70,5,70 | 50,5,50 |
| **Layer 7** | 70,3,70 | 70,3,70 | 50,3,50 |
| **FL** | 70,5,16 | 70,5,16 | 50,5,16 |

**FL :** flatten layer

Table 10 shows the classification results of the simple and combined MI EEGs with more than three classes. A different dataset was used for each study shown in Table 10. In the literature, it is observed that increasing the number of classes generally leads to a decrease in the classification performance. Likewise, the successes of simple and combined MI EEG tasks are observed to be low. In our study, the DivFE is used in the classification of simple and combined MI EEG signals. The use of DNN is one of the reasons why the performance is higher than other studies.

**Table 10.** The classification results for simple and combined MI EEG signals.

| Studies | type of MI EEG signals (number of classes) | mean accuracy % | Feature + classifier | number of subjects |
|---|---|---|---|---|
| Leon's study [36] | 3 simple + 1 combined | 51.6 | CSP + LDA | 6 |
| Yijie's study [38] | 3 simple + 1 combined | 54.2 | CSP + SVM | 8 |
| For MC2CMI in Leon's study [40] | 4 simple + 4 combined | 55 | CSP + LDA | 7 |
| Yi's study [41] | 4 simple + 3 combined | 70 | CSP + SVM | 10 |
| For III-IIIa in this study | 4 simple + 6 combined | **85** | only DivFE | 3 |
| For IV-IIa in this study | 4 simple + 6 combined | **78.6** | only DivFE | 9 |
| For datasets of [40] in this study | 4 simple + 4 combined | **77.8** | only DivFE | 7 |

## 4. DISCUSSION and CONCLUSIONS

The classification performance of the CNN based MI EEG signals for small datasets is generally increased by using transformation techniques such as the CSP [3, 4, 6, 7, 11, 14–19, 31], fast Fourier transform [5], short-time Fourier transform [9, 10], wavelet transform [12, 32] and the other methods [8, 13]. The success rates for two-class classifications are observed to be higher than that of a four-class EEG problem. Hence the transformation process becomes a must for the classification of four-class problems for boosting the performance. The CSP is the most utilized raw data transformation technique when the classification of MI EEG signals is of concern. The CSP and similar techniques are preprocessing steps which are used as a guide for the feature extraction phase of DNNs. Especially, the *m* parameter of the CSP affects the performance of the transformation and also the overall classification performance [15, 16].

In order to achieve high classification performances using a relatively small dataset, researchers have employed transformation techniques such as the CSP to transform the raw data but such techniques require higher computational power and thus become impractical for real life applications. In a previous study [33], researchers have investigated the effect of the augmentation process on the classification performance of MI EEG signals instead of using a preceding transformation such as the CSP. It has been demonstrated that by resulting in high success rates for the classification of MI EEGs, the augmentation process was able to compete with the CSP. In this study, there is no need to apply the augmentation process to the training set. The dataset automatically grows while generating combined EEG signals. For instance, in order to generate combined MI signals corresponding to the left hand-right hand, the left hand MI signal is simply added to the right hand MI signal. In this way, the number of combined samples is equal to the number ($M$) of the left (or right) hand samples. Hence, the number of samples in the combined MI EEG dataset can increase up to $M^2$.

To the best of our knowledge, the CSP was used as a preprocessing step in all studies that classify simple and combined MI EEG signals [36–41]. In the studies [36, 38, 40], it is observed that the CSP is used with the one-versus-rest method to increase the performance of the CSP. In fact, the use of CSP with one-versus-rest strategy in the classification of simple MI EEG signals increases the classification performance. However, this strategy did not improve the classification performance when using combined MI EEG signals [36, 38, 40]. The sources of the simple MI EEG signals are located in different locations of the brain, and each simple MI EEG signal consists of only one source. In this case, CSP easily reveals each simple MI EEG signal. In the case of using combined MI EEG signals, each combined

MI EEG signal consists of two sources. The low classification performances in the studies [36, 38, 40] were probably due to the use of the one-versus-rest strategy. In [41], a relatively higher classification performance was obtained, because the CSP was used in a multi-class strategy.

Due to the difficulty of acquiring EEG signals in real time, studies are generally performed with small data sets. In this context, it has been observed that studies have been conducted on artificially generated EEG datasets in recent years [42–44]. Among the generative techniques, the generative adversarial networks (GANs) with successful applications in image processing have gained significant attention [42]. This ability of GANs can be very useful for BCI systems where collecting large number of samples can be expensive and time-consuming. In [43], the BCI Competition dataset IIb was used in the training of GAN. It is observed that GANs can capture important characteristics of MI EEG signals, such as variations of power in the beta-band. Researchers were stated that the success of classifying the MI EEG signals is related to the size of the dataset used. In this context, the training set was enriched by generating artificial MI EEG signals to save time. It is observed that it is possible to replace up to 50% of frames with artificial data; and success results are obtained using a frame collection with 87.5% artificial frames.

The considered approach is based upon the assumption that the brain activity induced by the MI of the combination of both hands corresponds to the superposition of the activity generated during simple hand MIs. In simpler terms, this implies that the activity elicited by the combination of both hands MI can be considered as the superposition of the activity generated by simple hand MIs [36]. In the study [38], the obtained results showed that this simplification is convenient, and they confirmed that characterizing a multi-label task as the superposition of the involved sources represents a plausible model. To this end, they made the assumption that combined MIs can be modeled as the superposition of the activity generated by each one of the involved body parts. Also, in the study [41], researchers imply that there exist the separable differences between simple limb MI and combined limb MI, which can be utilized to build a multimodal classification paradigm in MI based BCI systems. The conjecture for artificial generation of combined MI EEG signals is validated on a dataset [40], and high classification results are archived by using artificially combined MI EEG signals in the training set. The main reason why the performance is very high compared to the results in [40] is the use of DNN and especially DivFE for this database. DivFE has previously been shown to have high successes in classifying MI EEG signals [33, 34]. When the DNN starts learning the training set by more than 90%, the accuracy for the test set continues to increase; however, it was observed that there was a decrease in the sensitivity. Therefore, the training process should be stopped to prevent over-fitting when the training accuracy rises above

90%.

It is observed in [36] that the artificially generated combined MI signal has the same characteristics with the combined MI signal acquired from the subjects. So, this artificially generated MI signal can be used as if it were a real signal acquired from a subject. Actually, artificially generated signal shows the ideal situation. However, it is possible that ERD signals do not occur at the same time like the artificially generated signals. The magnitude of the ERD/ERS signal decreases because the instant mean values of the two simple MI signals are added together in the generation process of the combined MI signals. This can be a problem with conventional classifiers. The fact that the two ERD signals that occur at different times in an epoch may not present a disadvantage for the DNNs.

It is well known that some subjects have difficulties in generating even a simple distinguishable motor imagery signal. Therefore, it will be even more difficult to generate combined motor imageries for those subjects. The use of the combined MI signals that is introduced in this paper can provide feedback for these subjects during the experiments. The experimental setup can be designed as follows. In the beginning, the subjects are trained with four simple MI tasks. Then, combined MI signals are artificially generated. In the third phase, DNNs are trained with a dataset created from simple and combined MI signals. Finally, these trained DNNs are used for feedback purposes so that the subjects can generate their own combined MI signals. After the temporary training phase is complete, subjects can be ready to create the datasets.

**Acknowledgement**

The work in the article is supported by the Istanbul Technical University Scientific Research Project Unit [ITU-BAP MYL-2018-41621 and ITU-BAP MYL-2019-41895].

**Compliance with ethical standards**

Conflict of interest: The authors declare that there is no conflict of interests regarding the publication of this paper.